\documentclass[journal]{IEEEtran}
\ifCLASSINFOpdf
\else
\fi
\usepackage[T1]{fontenc}
\usepackage[table]{xcolor}
\usepackage{graphicx}
\usepackage[cmex10]{amsmath}
\usepackage{amsfonts}
\usepackage{xcolor}
\usepackage{amsfonts}
\usepackage{amssymb}
\usepackage{amsthm}
\usepackage{booktabs}
\usepackage{algorithm}
\usepackage{algorithmic}
\usepackage[misc]{ifsym}
\usepackage{tabularray}

\usepackage{times}

\usepackage{soul}
\usepackage{dblfloatfix}    
\usepackage{url}
\usepackage{xcolor}
\usepackage{float}
\usepackage{cite}
\usepackage{multirow}
\usepackage{array}
\newcolumntype{p}[1]{>{\centering\arraybackslash}p{#1}}
\usepackage[pagebackref=false,breaklinks=true,colorlinks,bookmarks=false]{hyperref}

\usepackage[caption=true,font=normalsize,labelfont=sf,textfont=sf]{subfig}
\begin{document}
\hypersetup{
  pdfinfo={
    Title={Socio-economic landscape of digital transformation and public NLP systems: A critical review},
    Author={Satyam Mohla, Anupam Guha},
    Subject={},
    Keywords={}
  }
}
%
\title{Socio-economic landscape of digital transformation \& public NLP systems: A critical review}

%
%
%

\author{Satyam~Mohla$^{\dagger}$\href{https://orcid.org/0000-0002-5400-1127}{\protect\includegraphics[scale=0.12]{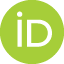}}~\IEEEmembership{Member,~IEEE,} Anupam~Guha 
\thanks{$\dagger$ Corresponding Author  \{\tt\href{mailto:satyammohla@gmail.com}{satyammohla@gmail.com}\}} 
\thanks{Satyam Mohla is currently at User Experience \& Data Solution Div, Honda Innovation Lab, Tokyo and during the course of this work, was affiliated with IIT Bombay, India \& Digital Asia Foundation. This work was supported in part by Moral Morpheus initiative at Digital Asia Foundation \& by UNESCO-LSUE Ethics in a Global Context Fellowship, University of Lucerne.}
\thanks{Anupam Guha is at IIT Bombay, India.}%

}
\maketitle

\begin{abstract}
The current wave of digital transformation has spurred digitisation reforms and has led to prodigious development of AI \& NLP systems, with several of them entering the public domain. There is a perception that these systems have a non trivial impact on society but there is a dearth of literature in critical AI on what are the kinds of these systems and how do they operate. This paper constructs a broad taxonomy of NLP systems which impact or are impacted by the ``public'' and provides a concrete analyses via various instrumental and normative lenses on the socio-technical nature of these systems. This paper categorises thirty examples of these systems into seven families, namely; finance, customer service, policy making, education, healthcare, law, and security, based on their public use cases. It then critically analyses these applications, first the priors and assumptions they are based on, then their mechanisms, possible methods of data collection, the models and error functions used, etc. This paper further delves into exploring the socio-economic and political contexts in which these families of systems are generally used and their potential impact on the same, and the function creep of these systems. It provides commentary on the potential long-term downstream impact of these systems on communities which use them. Aside from providing a birds eye view of what exists, our in depth analysis provides insights on what is lacking in the current discourse on NLP in particular and critical AI in general, proposes additions to the current framework of analysis, provides recommendations, future research direction, and highlights the need to importance of exploring the social in this socio-technical system. 
\end{abstract}
\vspace{-5pt}

%
%

\maketitle
\section{Introduction}

\IEEEPARstart{I}{n} the current landscape of digital transformation, Natural Language Processing (NLP) has risen to one of the most important sub areas of AI/ML research and has a significant impact on not just the computer sciences, but linguistics, social sciences, and economics as well, the last due to its now influence on how labour is done. Communication via language is one of the most fundamental trait of human communities throughout human historical development\cite{hurford1999evolution} and how language is used has always impacted socio-economic relationships in all human societies. Machines with the ability to use and generate human language (or “natural language” as it is called in the field) are changing and may further change multiple social and economic activities which were only performed by humans until recently. From translation of texts\cite{koehn2009statistical} to copy editing to interactive conversations, question answering\cite{mccann2018natural}, and eventually multi-modal fluent conversation\cite{daniel2020xatkit} the current ability and future goals of NLP are having and will have a significant impact on how humans work, how they interact with machine intelligent agents and eventually how they interact with each other. This may also influence and alter extant power structures of human societies. 


There is a growing realisation in the field of the ethical AI to decide what kind of NLP systems ought to be researched and which should be completely avoided\cite{leins2020give, bender2021dangers}. There is also an increasing realisation that these decisions are not to be relegated to the dimension of only ethics\cite{bietti2020ethics} as there is a long tradition of scholarship which posits that social, economic, and political relations which decide the arc of scientific progress\cite{borras2020roles} are not captured by the ethical lens. It must be kept in mind that alterations to human labour, society, economics, and politics etc. are not deterministic changes\cite{durham2019you} due to technological progress (in this case in NLP) as technological progress and framing of research problems does not happen in a vacuum, instead technological development including automation is a result of a series of political and economic choices by individuals, contending classes\cite{david2017forces}, and communities with differing interests. As such, this paper will attempt to not make an external trans-historical/moral critique of NLP but rather an ``immanent critique''\cite{antonio1981immanent} which broadly reviews the world of NLP research and then examines its impact while proposing additions and alterations to the current framework of how AI in general and NLP in particular is evaluated. 

The first contribution of this paper is a broad review into what are the kind of areas where SoTA NLP research has impacted the ``real world'' in the last five years. Seven such domains are examined and illustrated by 30 specific works. This is useful for anyone who wishes for a bird's eye view of research in applied NLP. The second contribution of this paper is to then analyse these seven interconnected families of research via instrumental, normative, and social lenses. The paper thus examines how is NLP broadly researched, designed, and deployed, and for whose benefits the field operates and also in a limited fashion examines on how that research gets impetus in the first place.The paper attempts to frame a future direction of research to measure long term impact of these NLP systems. We are aware that in doing so the paper itself alters the field of NLP ethics because it is making a decision on what kind of NLP systems deserve examination, commentary, and criticism, that is a weakness we acknowledge and hope our suggested future work complements.

\section{Related Works}
While the field of AI ethics has matured over the last few years, NLP specifically has lacked as much as of a specific analysis, and yet there has been real progress in some areas. First, bias in both datasets and (generally deep learning) systems have been widely investigated and acknowledged\cite{garrido2021survey} and multiple proposals like data statements and data sheets\cite{bender2018data, gebru2021datasheets, schick2021self} have been proposed to correct for bias in NLP datasets. There has also been an acknowledgement on the community's part of dual-use\cite{leins2020give} when OpenAI's GPT-2 was not initially released as a model over fears of abuse but finally was released after lack of evidence thereof and pressure from the research community. Thirdly, there has been attention drawn towards the real-world dangers of large scale use of language models, imputing meaning to what has been generated where none exists, and the connected financial and ecological impact of these systems\cite{bender2021dangers}. There has been, in the area of computer vision, criticism of the pseudo-scientific nature of affective computing\cite{stark2021physiognomic}, both in terms of trying to measure psychological phenomenon from external physiognomic cues and also these arbitrary systems being used to police people\cite{marda2021emotional} and perpetuate bias. We do a similar critique of affective computing in the context of NLP. Finally, new emerging works \cite{shmueli2021beyond, mohla2021material, kummerfeld2021quantifying} examine the ethics of crowd-workers used in NLP research and make a case as to why ``fair pay'' only scratches the surface of what the community needs to consider while using these platforms to collect data. This last work opens the door to the least talked aspect of NLP research, the consideration of erased labour which creates the data NLP learns on\cite{gray2019ghost}. These directions of analysis inform our work which is an attempt at gauging the landscape of what NLP systems exist in the ``real world'' and what are their broad impacts, social and economic. We hope our review work of a broad array of such systems contributes to the field of critical AI as well as informs NLP researchers.

\section{Systems being reviewed}
{Our methodology consisted of two stages. First, to develop an analytical framework to identify avenues of socio-economic analysis  missing from current discourse, a review was conducted. These avenues, namely instrumental analysis, normative analysis, function creep, political-economic impact \& long term impact are discussed further in Section 4. 

Second, a scoping review of documents, webpages and grey literature was conducted to capture ensuing changes across domains. We consider thirty such examples of NLP systems across seven domains. For each domains we will present these representative examples of work in the last five years which already have been deployed or have the potential to be deployed outside of the lab.}

\begin{figure}[h]
   \centering
   \includegraphics[width=\linewidth]{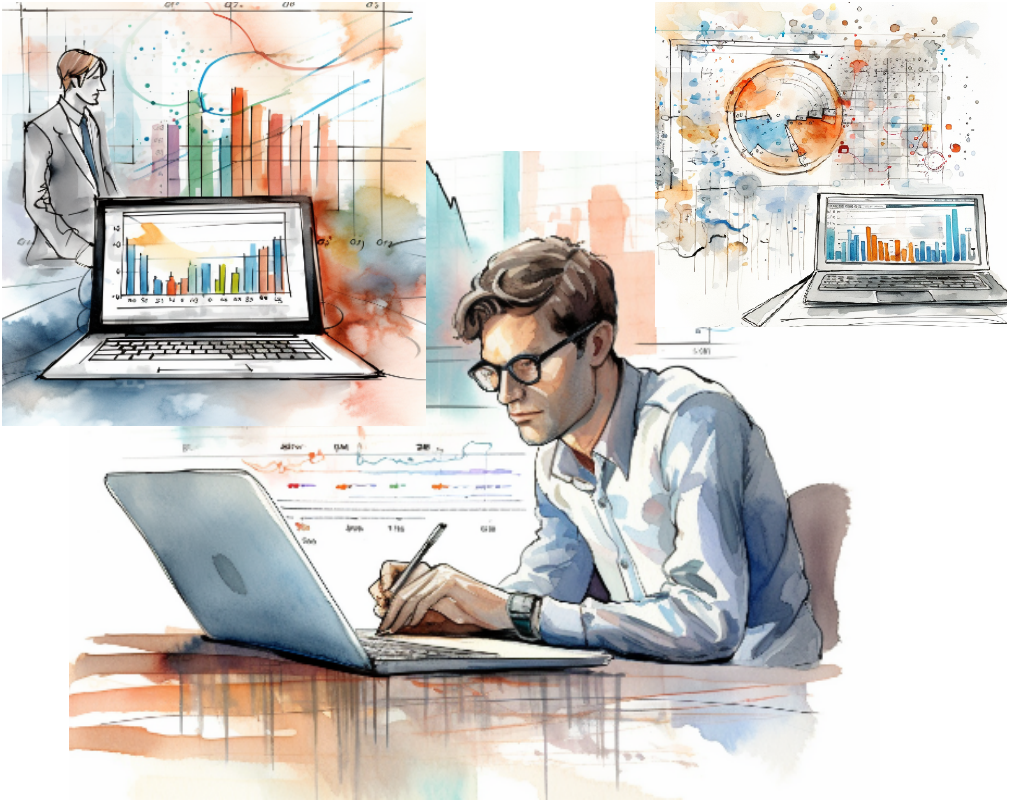}
   \caption{NLP Systems in financial decision making : Judging the creditworthiness and likelihood of defaults etc.}
\end{figure}

\subsection{Financial decision making}
The first family of NLP systems center around those applications aiming to expand the accessibility of finance by “judging” individuals. A set of problems in this family is concerned with the evaluation of creditworthiness, which complements and augments the traditional systems which use statistical methods on financial information with the use of machine learning methods. Another set of problems in this family is the use of NLP to identify who is likely to default on loans, essentially judging the trustworthiness of clients, and the third set of problems use similar NLP methods for various applications in the insurance industry, in the areas of marketing, underwriting, claims, reserving, and preserving. Illustrative examples in this family are:

1. SCOR SE reinsurance employs NLP techniques across various applications within this category\cite{scor}. Their strategic influence on insurance policy-making and company strategy is achieved through the application of NLP for analyzing commentary on social media platforms like Twitter and Reddit. Additionally, ongoing efforts are focused on utilizing NLP to extract valuable information from medical reports, aiding underwriters in addressing complex cases. SCOR SE is also actively involved in the development of NLP tools for claims analysis and fraud detection. Their aim is to anticipate claim developments and estimate associated costs through the utilization of these techniques. 

2. In their research, Netzer et al\cite{netzer2019words} present NLP techniques that go beyond the conventional reliance on financial and demographic information for assessing loan applicants which explores the potential of using textual information provided by applicants to predict their likelihood of defaulting. The study also explores the identification of borrower traits and investigates the correlation between defaulting loan requests and the writing styles associated with extroverted personalities and individuals exhibiting deceptive tendencies.

3. Crouspeyre et al\cite{crouspeyre2019creditworthiness}, advocate for the integration of NLP techniques to enhance existing creditworthiness measurement methods, which primarily rely on financial information such as FICO scores. Their proposed approach involves utilizing machine learning techniques and incorporating non-financial information, including phone log analysis and social media analysis. They claim that NLP techniques offer a less intrusive alternative, enabling the inclusion of individuals without a banking history. Additionally, they suggest that NLP techniques can effectively measure fraud by detecting inconsistencies or evaluating a borrower's business knowledge, among other factors.

\subsection{Customer Service}
The second category of NLP systems focuses on their application within the customer service industry, aiming to automate customer service processes and analyze customer feedback. These technologies encompass various aspects, including sentiment analysis to evaluate the handling of customer service calls, NLP techniques to analyze extensive text data from social media and online forums, and the development of NLP systems for creating interactive bots and virtual agents to navigate complex customer interactions. Below are illustrative examples within this category:

4. Jia\cite{jia2020deep}, conducts sentiment analysis on both textual and acoustic data obtained from a dataset of customer service calls. The dataset is annotated with positive and negative sentiment, and the study aims to incorporate multimodal features such as text and audio. The researchers explore a semi-supervised approach for annotation, enabling efficient annotation of a substantial amount of data. They evaluate multiple models to identify the most effective signal classification based on their feature set. The authors propose assigning a numerical sentiment score to each call using these features and models. 

5. Vermeer et al\cite{vermeer2019seeing}, attempts to investigate the efficacy of various NLP techniques, including sentiment analysis, dictionary-based methods, and supervised machine learning approaches, in categorizing Electronic Word of Mouth (eWOM) or online chatter about companies on social media platforms and online forums. Their objective is to prioritize relevant online conversations that companies should respond to. The study finds that content-based machine learning methods outperform sentiment analysis-based methods in identifying the most pertinent online chatter for companies.

6. The Government Technology Agency of Singapore has been actively experimenting with virtual assistants or chatbots \cite{governmenttechnologyagency_2022}. They claim that the implementation of chatbots has significantly reduced query wait times, increased accessibility, and improved overall user experience for citizens and businesses. Singapore, along with other countries, has witnessed the advancement of chatbot-mediated public services. For example, DigiMo by Niculescu et al\cite{niculescu2020digimo}, utilizes sequence-to-sequence deep neural networks trained on dialogue data specific to Singaporean citizens, considering emotional content within the data.

\vspace{-10pt}

\begin{figure}[h]
   \centering
   \includegraphics[width=\linewidth]{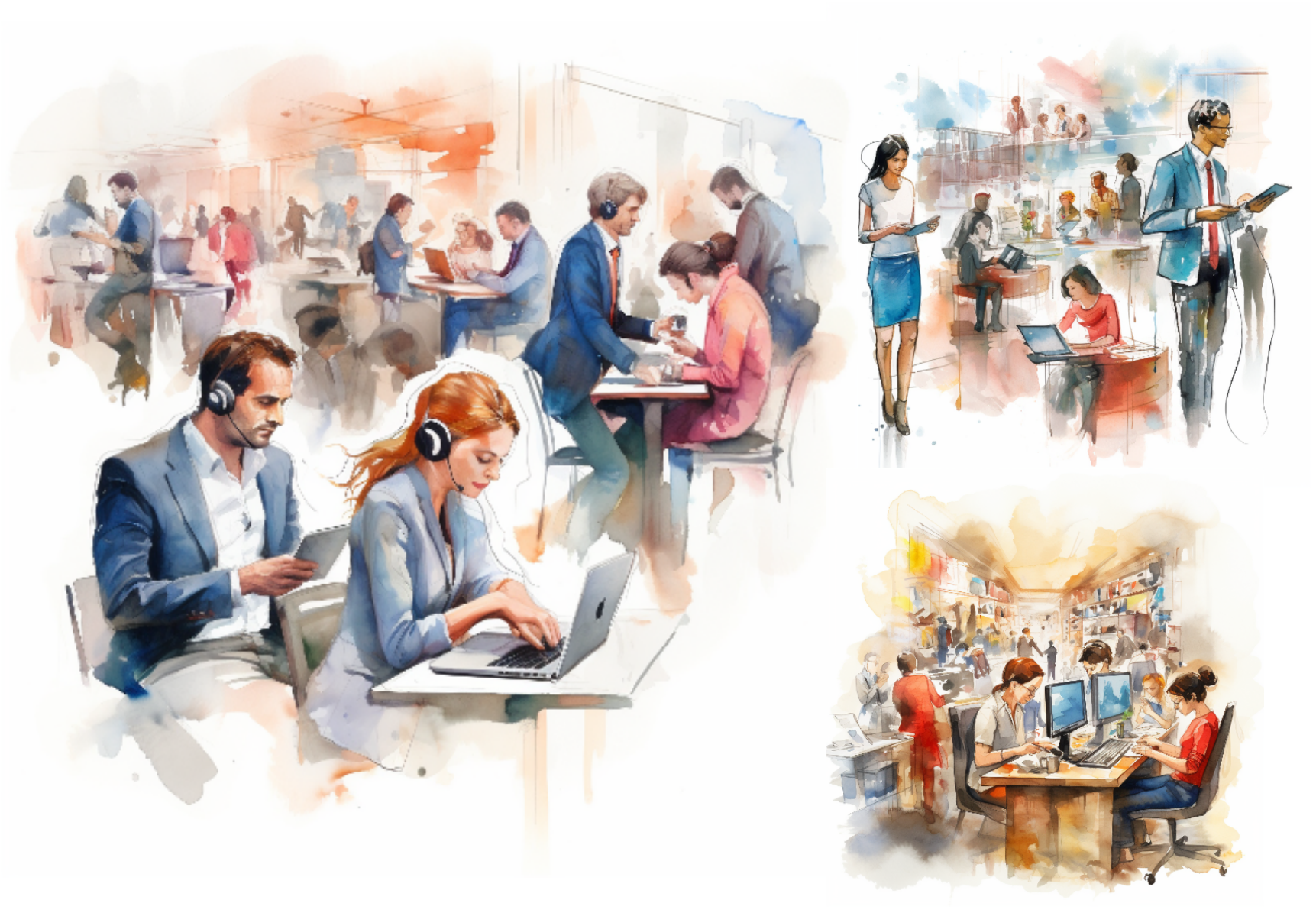}
   \caption{NLP Systems in customer service: Automating customer service, chatbots, feedback analysis etc.}
\end{figure}

\subsection{Policymaking and State}

\begin{figure}[hb]
   \centering
   \includegraphics[width=\linewidth]{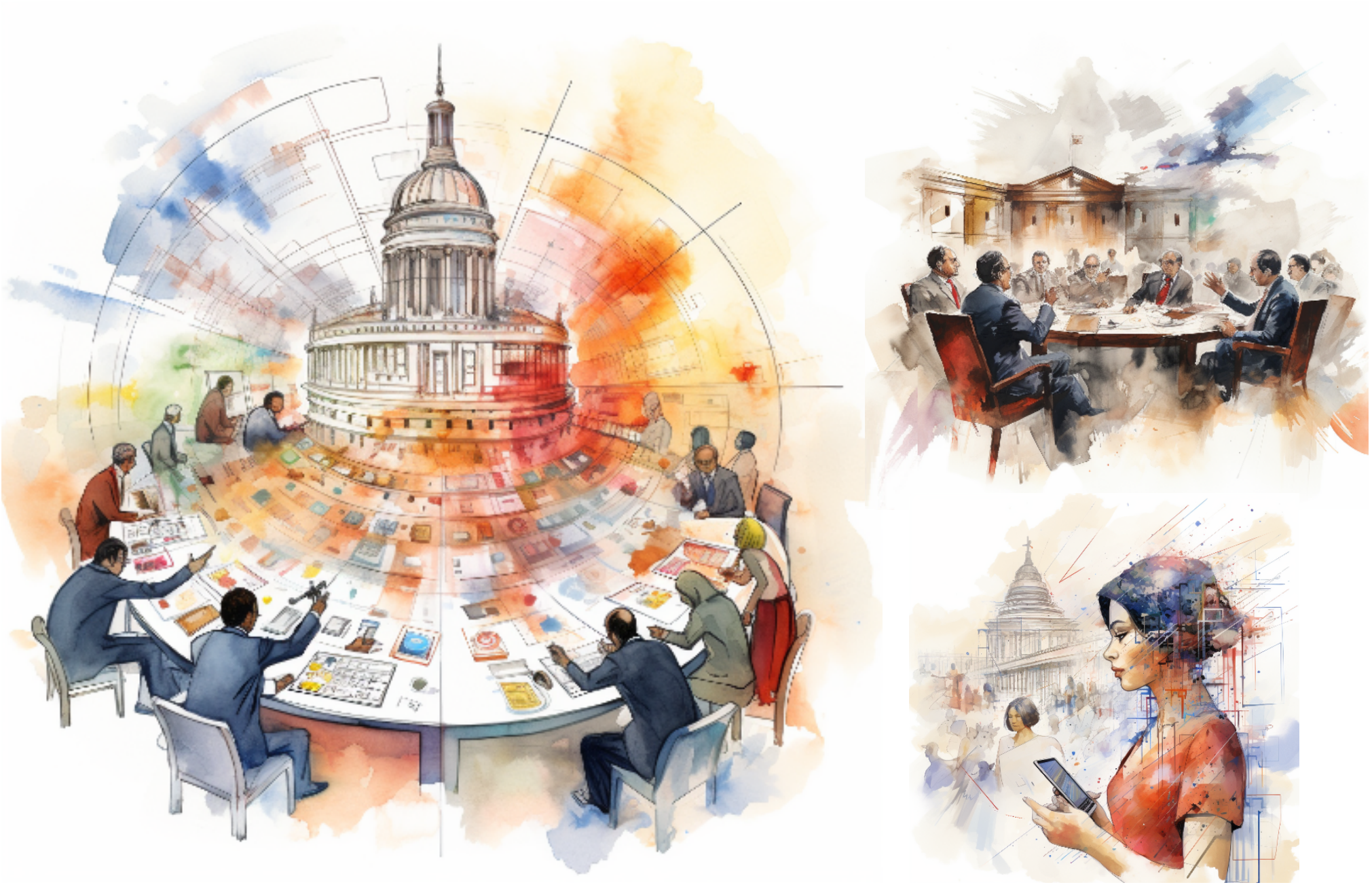}
   \caption{\mbox{NLP Systems in Policymaking and State: Public opinion} \mbox{analysis, sentiment analysis, evidence-based decision making etc.}}
\end{figure}

The third category of NLP systems currently under investigation focuses on their role in governance, aiming to inform and shape policy interventions across diverse areas. These NLP systems play a crucial role in evaluating large text corpora, ranking innovation mechanisms, investigating media texts to identify incidents on a global scale for detecting emergency triggers, evaluating public sentiment through text analysis, and aiding policy-making in "smart cities" by selecting relevant texts and organizing knowledge bases from vast scientific and technical document repositories. This section highlights notable examples within this category:

7. Lin et al \cite{lin2016ecosystem}, affiliated with the US Department of Energy's Oak Ridge National Laboratory, have developed a methodology that leverages NLP to transform text and numeric data into geographic mappings, enabling the detection and characterization of clean energy innovation ecosystems. By utilizing this tool, which detects, measures, and characterizes clean energy innovations, policymakers gain valuable insights into regional efforts, aiding in policy decisions and facilitating an understanding of the existing work in specific areas.

8. Xiong et al\cite{xiong2020digital}, present a case study on the use of social media posts to mine public opinion for informing crisis management policies during emerging environmental threats. They employ Latent Dirichlet Allocation (LDA) to identify topics discussed on Twitter related to the 2019 Chennai water crisis in India. The researchers classify tweets on this topic and investigate the relationship between social media chatter and rainfall precipitation levels. By analyzing these relationships, policymakers can gain insights into public sentiment and its correlation with environmental conditions, informing effective crisis management strategies.

9. Alam et al\cite{alam2020social}, propose a novel method for social media sentiment analysis specifically designed for smart city applications. They explore multiple hyperparameter combinations for neural networks to determine the most effective models for sentiment analysis using Twitter datasets. By interpreting social media sentiment, policymakers can gain valuable insights into public opinion and perception, supporting decision-making processes in the context of smart cities.

10. Pérez-Fernández et al\cite{perez2019corpus}, developed CorpusViewer, a powerful tool designed to support policymakers in analyzing a wide range of documents, including patents, scientific publications, and public aids, to gather evidence for policy implementation. CorpusViewer offers several automated functionalities, including document classification into a taxonomy, basic topic modeling, document similarity and plagiarism detection, semantic area identification, temporal analysis, and other useful analysis tools. This tool enables policymakers to efficiently analyze and extract relevant information from diverse document sources, facilitating evidence-based decision-making processes.

\subsection{Education}
Building upon the preceding category of NLP systems, the fourth category of NLP systems explored in this paper focuses on their applications in various educational contexts. NLP systems find utility in language learning apps, Massive Open Online Courses (MOOCs), automated teaching processes, tracking language competence evolution, automated essay scoring, and critical evaluation of essays. These systems also contribute to accelerating education in specialized areas. Here are illustrative examples within this category:

11. Automated essay evaluation is a significant focus within this category. Typically, AES tasks encompass three types: regression, ranking, and classification. Regression tasks involve predicting essay scores based on given metrics, ranking tasks aim to assess the quality of essays in a comparative manner, and classification tasks involve categorizing essays into quality categories. In their work, Bhatt et al\cite{bhatt2020graph}, enhance the existing methods for automated essay evaluation using NLP techniques. Their approach primarily relies on semantic sentence similarity, complemented by rule-based grammar and consistency tests. The goal is to replicate the features considered by human graders, thereby improving the accuracy and effectiveness of automated essay evaluation. 

12. Miaschi et al\cite{miaschi2021nlp}, propose an NLP methodology that employs computational stylometry to monitor the development of written language competence among Italian L1 learners. Their study focuses on predicting the chronological order of two essays written by the learners at different time points. They explore the language phenomena that contribute to this prediction task and examine how these phenomena evolve over time. Notably, their investigation emphasizes features related to the form of the text rather than its content.

13. Duolingo has conducted extensive research in the field of NLP methods, particularly in the context of language learning. One of their notable initiatives is the Second Language Acquisition Modeling (SLAM) task\cite{goenawanduolingo}, which leverages a substantial dataset comprising beginner-level student data from three different exercises. The aim is to model the process of language acquisition and learning for students. Additionally, Duolingo has introduced the Shared Task on Simultaneous Translation and Paraphrase for Language Education (STAPLE)\cite{mayhew2020simultaneous}. This task involves generating five translations for a given sentence and evaluating them against a weighted set of references, using a weighted F1 measure. This approach differs from traditional machine translation tasks, which typically compare the results to a single reference.  

\begin{figure}[t]
   \centering
   \includegraphics[width=\linewidth]{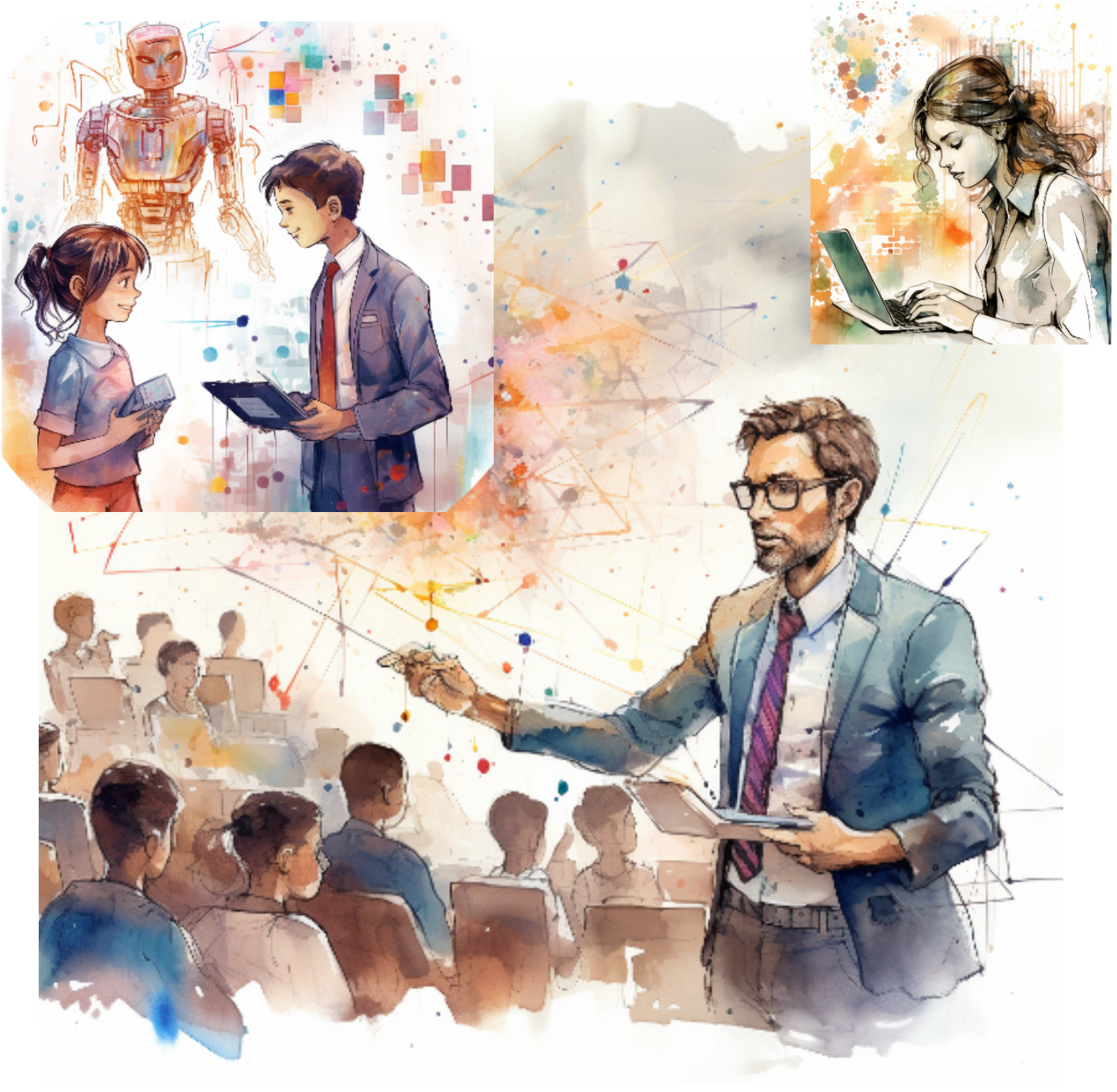}
   \caption{NLP Systems in Education: Automated assessment, essay scoring, language competence analysis etc.}
\end{figure}

14. With the rising popularity of Massive Open Online Courses (MOOCs), there is a growing need to automate certain aspects of the pedagogical process, considering the large number of students involved. One specific area of interest involves predicting instances where urgent instructor intervention is required based on student forum posts. To address this challenge, Alrajhi et al\cite{alrajhi2020multidimensional}, employ a combination of NLP methods and a deep learning model to effectively classify these forum posts and identify cases that necessitate immediate attention from instructors.

15. Poce et al\cite{poce2020correlations}, conducted research to identify the NLP features extracted from Italian text that provide insights into the critical thinking abilities of essay writers. They assessed critical thinking using six dimensions evaluated by human experts. The findings from this study, along with related works, contribute to the development of automated systems for evaluating critical thinking in student essays. By leveraging NLP techniques, these systems aim to provide valuable feedback on the critical thinking skills demonstrated in written compositions. 

\subsection{Healthcare}
The fifth area of exploration in NLP systems pertains to their application in medical and healthcare domains. These include the utilization of NLP by epidemiologists to identify and track the spread of infectious diseases through global media reports, the analysis of flight data to pinpoint disease outbreak locations, the generation of medical reports from patient-doctor conversations, the detection of diseases like early-stage Alzheimer's through text analysis, the acceleration of drug discovery through the review of scientific literature, the incorporation of NLP systems in telemedicine for automated interactions between doctors and patients via chatbots, and the provision of mental health advice through chatbot-mediated interviews. Illustrative examples in this category are:

16. BlueDot, a Canadian firm\cite{bogoch2016anticipating}, employs NLP to analyze a vast amount of textual data, including articles in more than 65 languages, global airline ticket data, censuses, climate data, public statistics reports, and global infectious disease alerts. By leveraging this data, BlueDot identifies disease hotspots and provides timely alerts to its clients. The effectiveness of BlueDot's automated infectious disease surveillance has been demonstrated during previous outbreaks such as the 2009 H1N1 influenza and the 2014 Ebola, and it accurately predicted the outbreak and high-risk cities for COVID-19 in 2020, nine days before official notices were issued by USCDC and WHO.

17. Enarvi et al\cite{enarvi2020generating}, explore NLP methods to automatically generate medical reports from transcripts of patient doctor conversation. They compare two methods, a hierarchical RNN with a pointer generator network and a transformer based sequence to sequence architecture. 

\begin{figure}[bh]
   \centering
   \includegraphics[width=\linewidth]{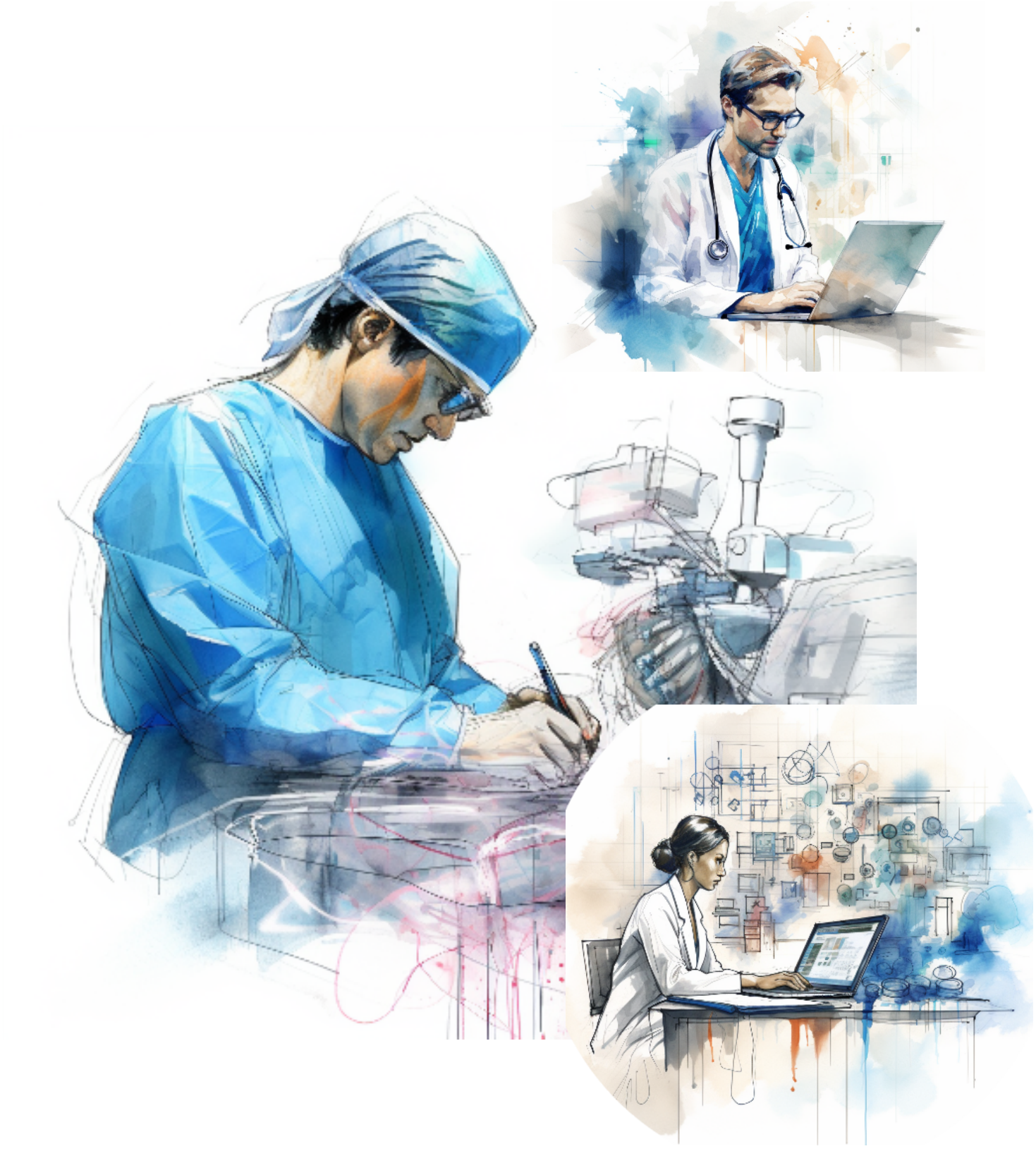}
   \caption{\mbox{NLP Systems in Healthcare: chatbots in mental healthcare,} \mbox{drug discovery, medical report generation, telemedicine etc.} \newline}
\end{figure}

18. Early diagnosis of Alzheimer's disease is crucial for effective treatment and management, as it accounts for ~60\% of the dementia cases. NLP techniques have proven valuable in detecting early symptoms like Mild Cognitive Impairment (MCI). Li et al\cite{li2020analysis}, created the B-SHARP dataset, consisting of speech transcripts on various topics, to aid in MCI detection using transformer encoders.

19. Drug discovery is a process which involves discovering protein targets using the principles of how certain compounds interact with protein. NLP methods offer the opportunity to automate this process by analyzing text-based representations of biochemical entities, which are readily available. Ozturk et al\cite{ozturk2020exploring}, investigate the use of NLP methods to analyze these text-based representations of chemical compounds, aiming to accelerate drug discovery.

20. The COVID-19 pandemic has seen a significant increase in the utilization of chatbots across various healthcare applications, including digital medical health. One notable use case is the application of chatbots in the field of mental health. Wysa \cite{inkster2018empathy},, a mental health chatbot with over 1.5 million users, employs Cognitive Behavioral Therapy (CBT) techniques to provide support. Another chatbot, Woebot \cite{fitzpatrick2017delivering}, also utilizes CBT methods and incorporates tailored surveys for a personalized user experience. Meadows et al\cite{meadows2020conversational}, conducted an analysis and comparison of these two systems, revealing that while Wysa tends to ask more open-ended questions due to the need for additional information, Woebot follows a more scripted approach. However, neither of these chatbot systems can replace the role of actual therapists.

\subsection{Law}
The sixth family of NLP systems the paper covers are those used in law.  In the legal domain, these systems are utilized for various purposes such as offering legal advice through chatbots, extracting pertinent information to support decision-making in a case, reviewing contracts to ensure completeness, conducting eDiscovery to assess the relevance of documents to a case, and generating legal documents for specific legal proceedings. Illustrative examples in this category are:

21. In their work on legal judgement prediction, Yang et al\cite{yang2020leniency} construct a dataset of the same kind of charge with trial information as well as information on the attitude of the suspect and create a model to discover the relationship between the suspects attitude and the penalty of the case demonstrating that there is a relationship between the two.

22. Ruggeri et al\cite{ruggeri2022detecting}, employ Memory-Augmented Neural Networks (MANNs) to detect unfair clauses in legal contracts and provide insightful explanations. Their approach involves training a MANN on a corpus of online Terms of Service, enabling the identification of unjust legal provisions and offering potential rationales for them. The authors evaluated multiple MANN configurations to enhance classification accuracy and explainability compared to previous methods, presenting a valuable dataset for further research.

23. Queudot et al\cite{queudot2020improving}, design two task-specific legal chatbots: one focused on providing immigration-related information using data from the Government of Canada, and the other dedicated to job-related legal inquiries for bank employees. These chatbots are trained on FAQ data, leveraging the RASA model. The first chatbot utilizes online data obtained from the Government of Canada's Immigration and Citizenship Help Desk, while the second utilizes a bank's internal FAQ.

24. Sugathadasa et al\cite{sugathadasa2018legal}, create a mechanism for legal document retrieval by integrating two document vector representation methods. Their work highlights the value of incorporating semantic similarity measures in information retrieval tasks involving domain-specific legal documents, offering improved retrieval performance.

25. Automatic summarization of legal documents has been an active research area. Jain et al\cite{jain2020fine} employ a Bayesian optimization approach to fine-tune the hyperparameters of the Textrank algorithm, a classical text summarization technique. By optimizing an objective function based on the ROUGE score, their approach enhances the summarization performance of Textrank for legal documents, demonstrating its efficacy in generating concise and informative summaries.

\begin{figure}[t]
   \centering
   \includegraphics[width=\linewidth]{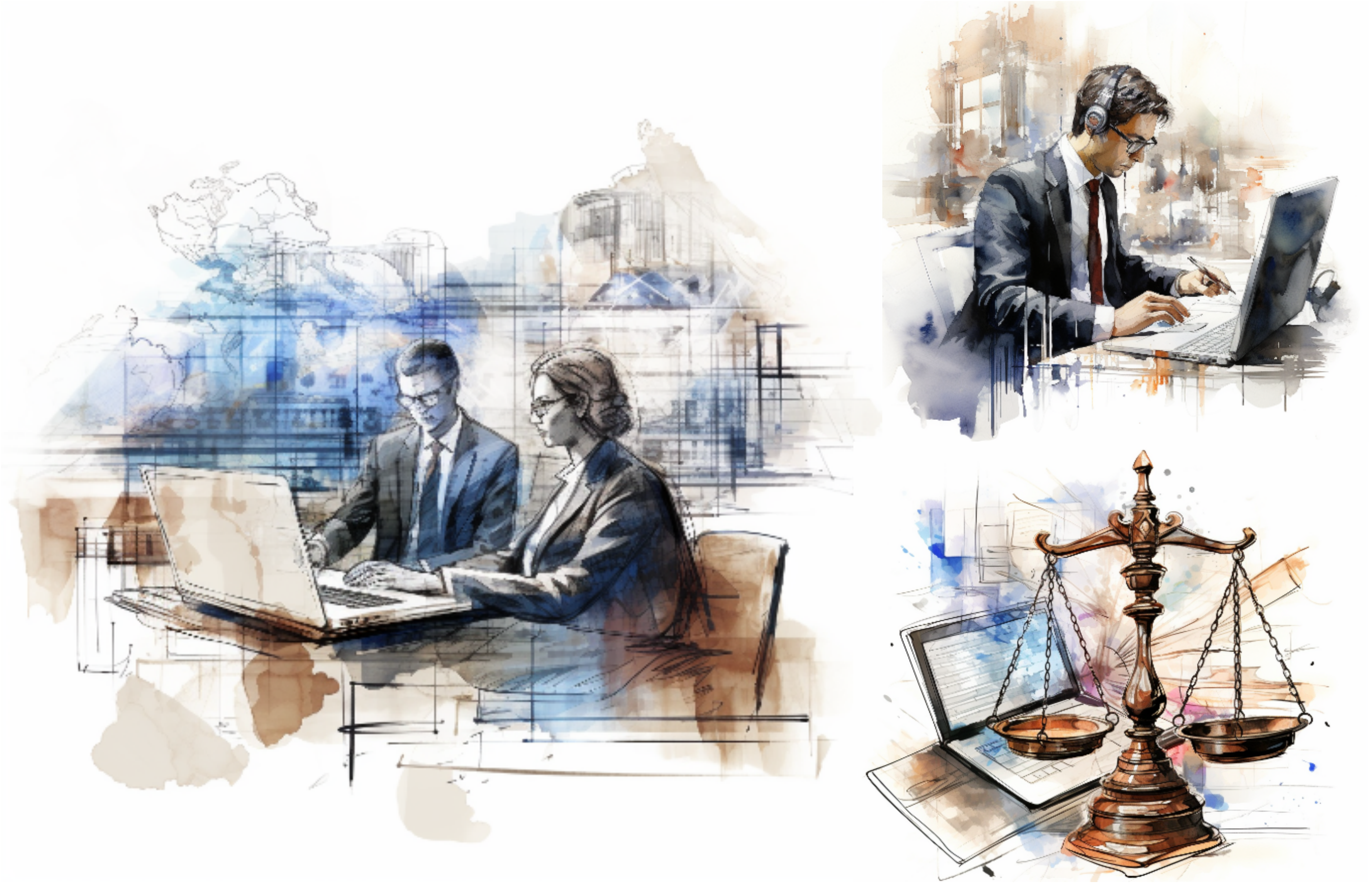}
   \caption{NLP Systems in Law : contracts analysis, immigration virtual assistant, legal document retrieval, summarization etc.\newline}
\end{figure}

\subsection{Security}
The seventh and final family of NLP systems discussed in this paper pertains to their application in law enforcement, surveillance, defense, and national security. In these domains, NLP systems play a crucial role in identifying hate speech, examining the radicalization of individuals and communities, and complementing other AI mechanisms in predictive policing. Additionally, NLP systems are utilized for automating surveillance of online communities and individuals, as well as refining metadata. Several illustrative examples in this category are:

26. Alshehri et al\cite{alshehri2020understanding}, developed a model to detect intentional threats, particularly dangerous speech, in Arabic Twitter. They constructed a dictionary of physical harm threats in Arabic dialects and collected a sizable dataset of threat-related information. By manually annotating a portion of their dataset, they analyzed the prevalent types of threats. The authors trained BERT variants using Arabic emotion data to identify these threatening texts accurately. 

27. Araque et al\cite{araque2022ensemble} leverage insights from affective computing, a field that focuses on computational methods for detecting and processing human emotions, along with SenticNet, a knowledge base for concept-level sentiment analysis. They employ two feature extraction methods to enhance their classification performance in hate speech detection tasks. Previously, Araque et al\cite{araque2022ensemble} classified radical text against neutral or anti-radical text using an emotion lexicon that annotates words with emotions, and a radical lexicon that measures the semantic similarity of a text using word embeddings. The authors found that incorporating emotion features improved the classification results.

\begin{figure}[t]
   \centering
   \includegraphics[width=\linewidth]{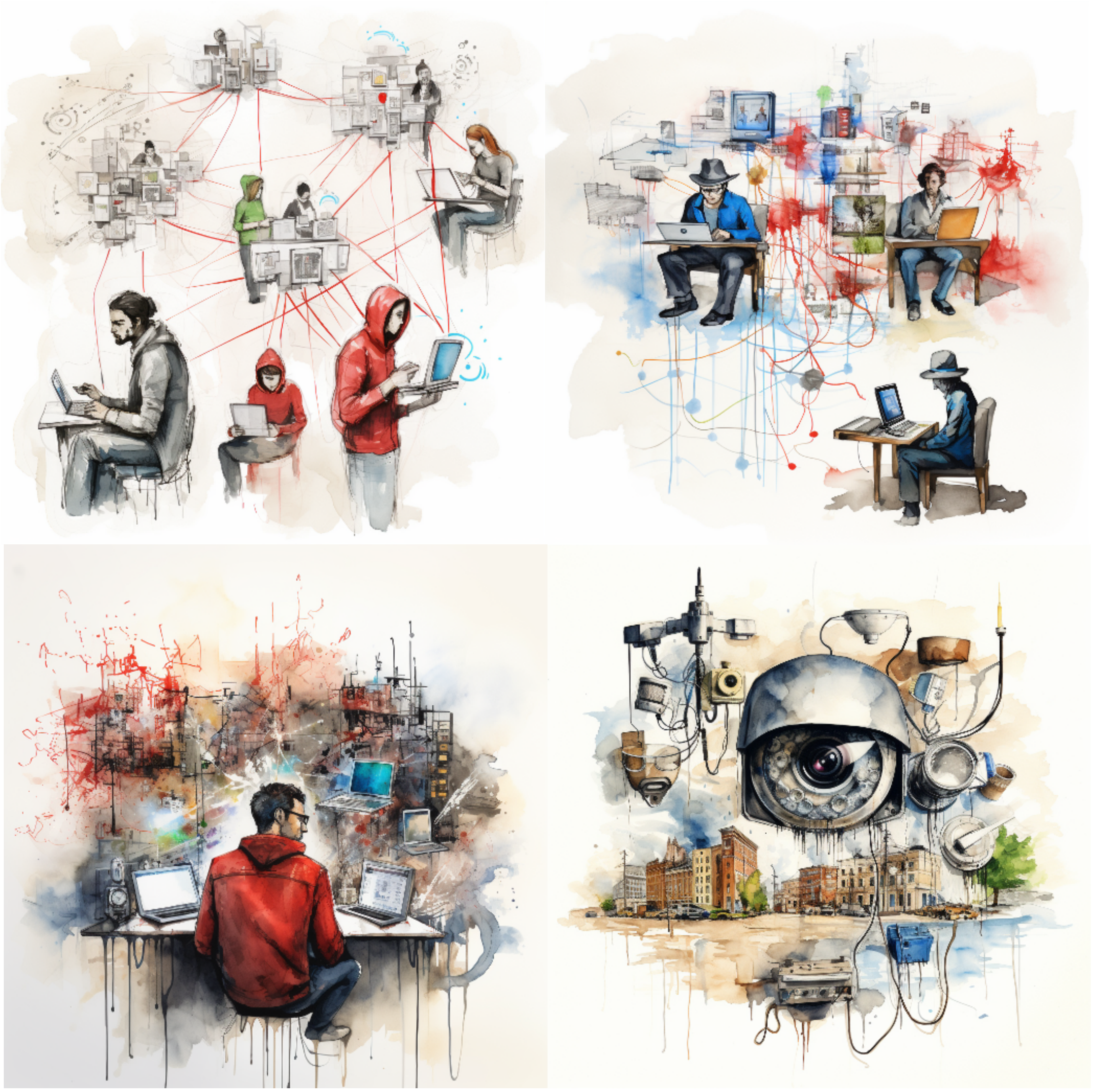}
   \caption{NLP Systems in Security : Surveillance, crime records analysis, digital forensics, cyber bullying etc.\newline}
\end{figure}

28. Percy et al. \cite{percy2018text} employed NLP tools to analyze crime records as text documents from diverse regions and time periods, aiming to predict crime patterns. Their approach involved identifying regions with similar crime patterns and clustering them over time.

29. Sun et al\cite{sun2021nlp} developed a digital forensic investigation platform using NLP techniques to identify criminal activity within online communities. The platform analyzed a corpus of communications, considering both the senders and receivers (or at least the sender in the case of social media). The method employed topic modeling to uncover the main discussion topics, refined those topics, and utilized a series of classifiers with the topics as features to identify criminal behavior among participants.

30. Ziems et al.\cite{ziems2020aggressive} conducted a study on cyberbullying detection, introducing a new annotation framework and creating a new dataset using Twitter data. Their objective was to distinguish cyberbullying from other forms of online aggressive language that existing classifiers cannot specifically identify.

\section{Analysis}
In this section, we will analyze and evaluate different aspects of these NLP system families, both their instrumental aspects and others, from various perspectives. While we acknowledge and address the potential weaknesses and unintended negative impacts that these systems or their variants may have, we do not disregard the intended benefits that motivated their creators. It is important to recognize that when these systems are used within their designated scope and constraints, the benefits typically outweigh the harm. The majority of harm arises from their misuse or deployment outside of the specific contexts they were designed for, which we will further explore in the subsection on function creep.

\subsection{Instrumental analysis}
\label{subsec:instan}

Evaluating NLP systems requires a comprehensive analysis of the underlying algorithms in their specific applications. This direct approach enables us to identify the strengths and weaknesses of NLP systems across various contexts and determine their effectiveness and limitations. Essential to this assessment is the careful examination of dataset collection methods and the inherent priors and assumptions embedded within them. By scrutinizing these methods, we can gauge the degree of fairness, accountability, and transparency (FAT) that these algorithms exhibit and uncover any potential biases that may arise. One may then evaluate the strengths and weaknesses of the specific models used in the aforementioned use cases. When tasks are well-defined, we gain valuable insights into how an NLP system may encounter failure points, thus enhancing our ability to predict and address potential shortcomings.

Consider NLP tasks that involve complex semantic analysis, such as the examples 2 and 3, which aim to establish connections between human tendencies and language usage. Building datasets for these tasks can be challenging due to the inherent difficulties in achieving annotation agreement. Annotators proficient in a particular language are typically employed to create annotated datasets for NLP tasks. They can effectively identify relevant aspects of text for tasks like coreference resolution or question answering. However, when it comes to tasks like irony/sarcasm detection, aggression analysis, assessing artistic merit in poetry, or identifying humor in jokes, reaching consensus becomes increasingly difficult. This is because the relationship between mental states or semantic content and their expression in natural language often lacks clarity or presents ambiguity. Furthermore, psychologists themselves have differing views on the existence of universally articulated emotions, adding complexity to mapping affective computing concepts to textual patterns. Some argue that emotions are universal aspects of human experience, while others contend that emotions are culturally specific and shaped by social and historical factors. This lack of consensus further complicates the process of creating datasets that accurately capture emotions in text, making it challenging to develop algorithms that can effectively recognize and respond to emotional content. At the extreme end, certain claims about mapping internal emotions to patterns of text may be unfalsifiable and fall into the realm of pseudoscience.

To illustrate this further, let's consider two cited NLP systems: system 4 claims to possess a deep understanding of sentiment, while systems 11 and 15 aim to evaluate the "quality" of written essays. However, these use cases inherently face the risk of arbitrariness, even among human evaluators. Compounding the challenge of arbitrary datasets and use cases is the probabilistic nature of off-the-shelf NLP models. Regardless of their performance, these models are still susceptible to errors. Additionally, many NLP models lack transparency, making it difficult to understand the reasons behind their errors. When these models fail, their incorrect results can be arbitrary and unforeseeable. This aspect becomes particularly crucial in domains such as policy, governance, justice, financial decisions, and security mechanisms, where arbitrariness is unacceptable. While ongoing research focuses on improving the explainability of NLP systems, the current standards fall short of what is necessary for human policymaking. Therefore, the use of NLP techniques in such applications should be cautiously employed, providing insights under strong checks and balances with human oversight. They should not be relied upon to make decisions autonomously. The literature reflects an ongoing debate regarding the limitations of accountability and transparency in these specific use cases.

In the case of systems 2 and 3, we observe the implementation of invasive decisions that assess customers' creditworthiness and predict future behavior, which can often be both erroneous and arbitrary. The involvement of algorithms in making these decisions creates a false sense of credibility, leading employees to rely on opaque AI systems without understanding the reasons behind customer rejections. Examples like systems 26, 27 and 30 pose multiple potential sources of error when applied in real-world use cases. Issues can arise from the data itself, incomplete annotations that fail to account for all contextual factors, overfitting of models to the data, and misinterpretation of the model's results. These systems are prone to breakdown when used in linguistic contexts outside their intended and tested domains. It is crucial to avoid using NLP tools for decision-making in domains that involve social and economic rights, as the decisions made by these tools can be incorrect, arbitrary, and lacking in transparency. While human agents in similar applications may also make mistakes or harbor biases, the legal and moral responsibilities of human agents can be addressed through existing laws and regulations, which do not apply to algorithms.

Another field where NLP techniques find application is in recommendation algorithms, which determine useful instances from a collection of data. These algorithms have benign applications in media websites, e-commerce, and policymaking (as seen in system 10). However, there is a recent trend of using them in law and judiciary contexts, like in systems 21,22,23 and 24. These techniques excel at accelerating the document discovery process, a laborious task in the legal domain, thereby reducing delays and increasing accessibility to legal resources, which are often limited for the general public in many countries. Another use case involves using NLP document retrieval to generate templates and offer automated legal advice, similar to system 6. However, there is a potential concern with recommendation systems, as documents deemed useless may never reach human observers for consideration. While this might not be problematic in media or e-commerce settings, in the legal environment, where missing documents could severely impact the life, rights, and liberty of individuals, the use of such services must allow for human-led discovery. In contexts where there is a lack of human resources for document discovery, but no immediate risk of harm from missing documents (e.g., scientific research, art, media, and scholarship), these methods prove extremely beneficial.

When NLP tools are used to provide insights rather than decisions, often unsupervised learning tools like clustering and topic modelling is used to group documents, discover structures and hierarchies to automatically organise them. Examples include systems 7, 8, 9, 10, which use variants of topic modelling and as long it is understood that such unsupervised techniques only display statistical aggregates and not specific answers, they can be safely used to get insights from. 

\begin{figure*}[t]
   \centering
   \includegraphics[width=\linewidth]{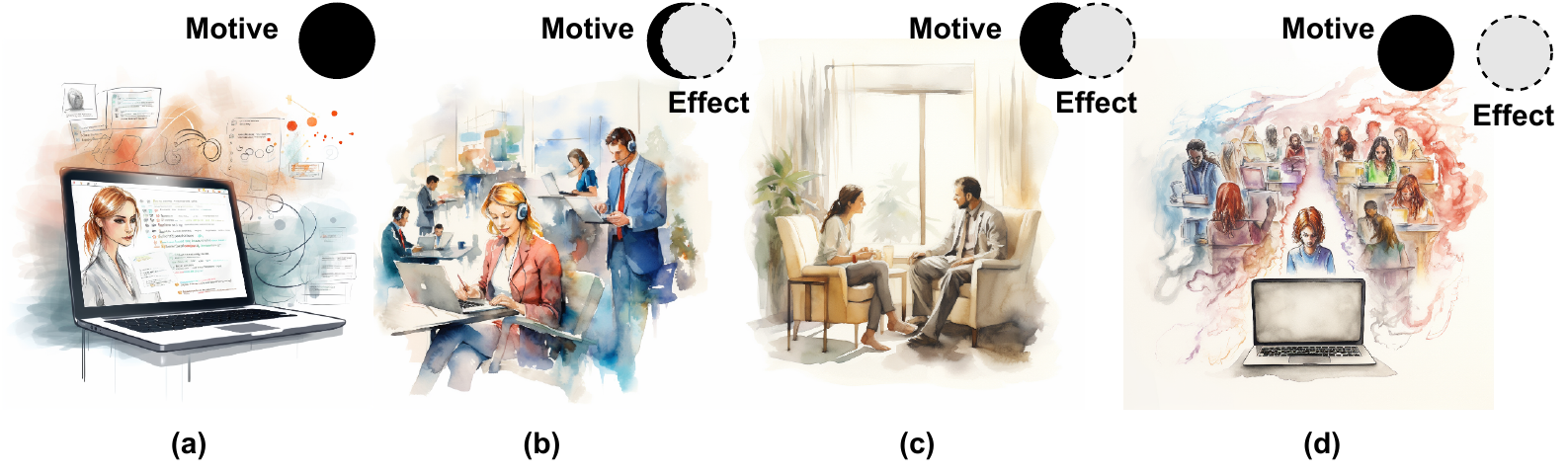}
   \caption{\textbf{Function creep can be understood as a divergence of motive and effect} (a) Example: Consider chatbots and virtual assistants (b) Being used by marketers and employers to improve the engagement and interaction with customers online. The motives and effect align significantly (c) Being used to augment or replace mental health professionals like psychologists, psychiatrists, therapists etc. and provide at-convenience therapy. The motives and effect have started to diverge. (d) Being used to spread misinformation or to fill chat rooms with spam and advertisements. The initial motives and resulting effect have significantly diverged.}
\end{figure*}

\subsection{Normative analysis}
The limitation of instrumental analysis, including the FAT framework, is its neglect of the ethical and human rights implications associated with NLP systems. This approach fails to consider the societal and political values that these systems may promote or hinder. For example, while system 11 may have technical flaws and dataset representation issues, the primary concern with it is that even with an excellent dataset
and tested algorithm, there is still potential for arbitrary exam results without human accountability. The use of such a system in critical decision-making, like determining exam results with significant consequences for students' futures, raises ethical concerns. The risk outweighs any benefits in terms of speed or scalability, warranting a careful examination of its usage. It is important to acknowledge that NLP systems are not socially and politically neutral when deployed in the real world. Some of these systems make implicit assumptions about desirability without explicit articulation, potentially perpetuating undesirable historical patterns. However, we must be cautious about overstating the case for normative analysis, as there is a lack of consensus on ethics within the AI ethics community, and accountability mechanisms for ethical violations remain insufficient. The discourse on ethics can be diluted by what some observers refer to as "ethics washing," where ethics is used as mere rhetoric without concrete commitments. To provide a more concrete ethical lens, grounding the analysis in human rights principles, as proposed by Marda\cite{daly2019artificial}, can offer a stronger foundation. Human rights possess a well-established enforcement framework and international recognition. Principles such as business and human rights emphasize the responsibility of states to safeguard human rights within their jurisdictions, even when non-state entities like businesses or developers are involved.

Among the systems listed, system 28 is one which predicts preponderance of crime in a region. As has been observed in the past, predictive policing has a habit of “predicting crime”
in overpoliced regions which are inhabited by the poor and racial minorities by replicating patterns of police behaviour. Relying solely on past patterns of crime carries the risk of over-determining future outcomes and influencing repressive state policies. In scenarios where these tools are often used without recognizing their inherent limitations, the ethical implications of such use should be questioned, as it attaches the label of criminality to the residences of an area and leads to collective guilt, and unethical authoritarian practices. System 20 uses chatbots to provide a semblance of mental health care. While chatbots explicitly disclaim their role as substitutes for human therapists, what is missing is the recognition of
the phenomenon that individuals may still attribute moral agency to these systems, as evidenced by past experiments such as ELIZA\cite{jain2018evaluating, weizenbaum1966eliza}. Delivering medical care solely through statistical algorithms is ethically questionable as it disregards the essential requirement of moral agency inherent in the practice of medicine. Systems 1, 2, and 3 extend the logic of existing credit scoring systems by incorporating NLP techniques to draw inferences about individuals' character. Apart from technical weaknesses related to arbitrariness and the impossibility of deterministic outcomes, these systems raise ethical concerns due to the potential economic harm they can inflict on individuals based on presumptions of intent rather than actual actions. This approach undermines fairness and can lead to unjust outcomes.

The last category, security, presents the highest potential for unethical use of NLP systems. This is primarily due to the exceptions granted to data security and human rights mechanisms within the carceral and security apparatus of various countries, as well as a history of problematic implementation and misuse of digital systems. For instance, systems 26 and 27, which should not be employed without human oversight, can be utilized by law enforcement agencies. However, there is a historical pattern of misuse and overreliance on machine learning systems, such as Facial Recognition Technology (FRT), which raises concerns about privacy and civil liberties. Even with human oversight, the presence of false positives remains a concern, as evidenced by attempts to introduce human moderators on various platforms to verify content flagged by ML algorithms as hate speech. However, the reliance on algorithms often leads to inadequate hiring of moderators, resulting in excessive workloads and detrimental effects on their mental well-being due to continuous exposure to violent content. In security settings, the insights provided by machines are often afforded more credibility than those provided by humans. This can lend legitimacy to clear violations of human rights and natural justice. The potential harm of these systems is further exacerbated by the gap in technical literacy between researchers and policymakers. This disconnect often leads to the systems being used in unintended ways, giving rise to function creep and straying from their original purpose.

\subsection{Function creep}
Function creep refers to the gradual expansion of the applications and uses of a particular entity beyond its original intended purpose. Initially, this term was coined to describe how data and datasets collected for ostensibly benign and beneficial reasons end up being utilized in multiple other areas without the knowledge or consent of the individuals whose data was originally collected. This phenomenon is prevalent in countries lacking comprehensive data protection legislation. In the context of NLP methods, we speculate on the potential for function creep based on past instances where it has occurred. A notable example is the case of COMPAS, an acronym for Correctional Offender Management Profiling for Alternative Sanctions\cite{brennan2007compas}, an algorithmic tool used in the US to predict criminal recidivism. Originally, COMPAS was designed to assess the specific needs of convicted inmates, such as mental health treatment, to facilitate their rehabilitation process. However, its application expanded to include decisions on matters like granting bail or determining the conditions for releasing convicts. This expansion occurred without sufficient recognition of the risks associated with granting algorithmic systems significant control over individuals' liberty. This is also despite the developers of the COMPAS system not initially designing it to be used for sentencing or being confident of its use in sentencing. Despite concerns raised about biases against racial minorities in systems like COMPAS, the fundamental question arises as to whether systems like COMPAS should be used at all, regardless of their biases, given that human rights scholars universally agree that decisions affecting individuals' right to liberty should not be based on assumptions about their future behavior.

Digital forensics datasets, as seen in systems 29 and 30, provide valuable resources for researching platform regulation and identifying patterns of hate speech and amplified threats. However, there is a risk of these tools being directly used by law enforcement who may not grasp the nuance that these systems need human verification, regardless of the margin of error.

Chatbot-mediated public services, exemplified by system 6, offer accessibility and speed in interactions. Nevertheless, they should not be utilized in situations where the absence of a human official could potentially harm a petitioner who requires discussion and assistance with complex problems. Similarly, in telemedicine, the use of chatbots, as depicted in system 20, must not undermine the responsibility of healthcare providers or create legal immunity. These applications should provide the option for interaction with a human professional to ensure the well-being of individuals. Topic models, such as systems 8, 9, and 10, are valuable tools for obtaining high-level insights to inform policymaking. However, it is essential to recognize that these insights are subject to interpretation and should not substitute the diligent collection of pertinent facts on the ground when making decisions. System 17 offers an automated summarization mechanism for medical interviews, which could play an expanded role in areas with a shortage of medical practitioners, and thus must never be used in applications in which missing out on some pertinent fact in a summary could result in actual harm to the patient. System 11, designed for automatic essay evaluation, has the potential to extend beyond controlled pedagogical spaces with human oversight and enter self-teaching apps. This could create a misleading perception of speed and convenience, compromising the true potential of effective teaching. Such harms are difficult to detect due to the misplaced trust people often have in technology. 

 The authors want to emphasise that the informed repurposement of technology for new use often inspires and drives innovation, and is not under scrutiny here. However, uninformed use deviating from its original purpose without understanding initial assumptions and systemic properties could lead to unintended consequences, including socio-economic impacts. Function creep often arises due to the profit motive of AI vendors, who may downplay the weaknesses of contemporary NLP systems, as well as the desire of businesses to avoid technology regulations. Additionally, it can stem from solutionism among state officials and policymakers who seek to avoid their responsibilities in the policy space. This necessitates a political-economic analysis of these systems to understand their broader implications.

\subsection{Political economic analysis}
Political economy examines the intricate relationship between the subject of study, in this case, different categories of NLP systems, and the interconnectedness of society and the state. Our primary focus is understanding how these NLP systems both impact and are influenced by socio-economic relationships, governments, and public policy. To begin, let's explore their effect on the mode of production, which refers to how a society as a whole produces goods and services. When analyzing these NLP systems from this perspective, we need to consider the labor involved in their creation and how they affect the workforce. Data production and/or collection play a significant role in the labor required for NLP systems across all categories. In certain categories like finance and customer service, data production involves activities such as data collection, cleaning, structuring, and annotation, which are carried out in-house by private companies or government organizations. However, in some instances, such as systems 8, 9, and 10, the data is obtained from the public domain through methods like scraping social media or accessing other repositories of public data. While regulations regarding the use of public data for commercial and policy-making purposes vary from country to country (for instance, GDPR in Europe prohibits certain data collections and uses), there is a broader debate surrounding "community data", which refers to data obtained from communities of people for the benefit of those communities. Some data, although publicly available, are not generated by individuals, such as weather data, and can be freely utilized. Another aspect of data-related labor in these systems arises when data is collected via crowdwork platforms like Amazon Turk or CrowdFlower. The proliferation of these projects contributes to the growth of platform work. Ongoing debates center around the ethical considerations of utilizing workers in these systems, as the work carried out on these platforms is generally not regulated by labor laws. Furthermore, these platforms often classify the individuals they employ as contractual associates rather than workers. This situation implies that jurisdictions with robust labor laws but lacking regulations on crowdwork platforms may unintentionally incentivize a shift towards certain NLP use cases. For example, this could involve reducing the reliance on medical practitioners and instead utilizing systems 17, 18, or 19 for tasks like manual drug discovery or promoting legal AI systems like example 23 instead of hiring more employees. The underlying argument is that these systems may not necessarily perform tasks faster or better. However, due to the "hidden" nature of their data collection and the potential exploitation of crowdworkers, they generate more profit and further promote platformization.

The utilization of NLP systems leads to a widespread state of job insecurity and decreased wages among workers in the respective field. Moreover, this precarious situation disproportionately affects marginalized communities, particularly women who face a dual exploitation in both their domestic and professional lives. Furthermore, employing these systems in critical areas such as healthcare and the judiciary may initially appear appealing from a short-sighted economic perspective. However, it is actually detrimental since these systems are inherently stochastic, prone to errors, and undermine official accountability as previously mentioned. As a result, the costs are shifted onto vulnerable members of society.

Regarding their influence on labor, certain systems in education and healthcare, such as automatic essay evaluation, NLP applications in language learning platforms, and NLP applications in MOOCs, are altering work dynamics in these fields. If these systems are perceived to reduce the costs associated with hiring and training teachers, they are often implemented without sufficient consideration for potential negative effects. This is already occurring in many countries, particularly in the global south. In the current global climate of privatization, where governments are retracting from their welfare responsibilities, private companies are assuming control over tasks such as education, healthcare, and even aspects of the judiciary and policymaking. These companies operate with distinct priorities compared to democratic governments and view NLP techniques as labor-saving measures that yield profitability. Concern arises because the changes in education precede the thorough examination by education experts and pedagogues to determine if they may have adverse effects on students. Often, the proliferation of these systems presents a fait accompli to policymakers before their impacts are fully analyzed.In healthcare, examples 17 and 20 illustrate potential bandaid measures that states and companies may adopt if they are perceived as sufficiently robust, with the aim of avoiding the expenses associated with training and hiring medical professionals. However, the actual impact of these stochastic (and thus unintelligent) systems is not thoroughly studied by the time public policy is replaced by NLP infrastructure developed and maintained by private companies.

Thus we observe that certain systems have a notable indirect impact not only on governance and policy-making but also on the very fabric of what constitutes governance and policy-making. The utilization of these systems by private companies has the potential to reshape the landscape, gradually eroding the space for policy deliberation and replacing it with a mechanism that is inherently stochastic and lacks accountability, as machine learning itself is a stochastic process and only humans possess moral agency. Furthermore, the sections discussing policy-making and legal tools, including systems 7, 8, 9, 22, 24, and 25, demonstrate that readily available systems are actively altering the functioning of these domains. For instance, tools like system 24 and 25 have the capacity to disrupt the way burden of proof operates, thereby diminishing accountability by obscuring the relevance of specific documents and their textual content. Similarly, systems like 8 and 9, which rely on highly stochastic techniques to inform decision-making, introduce greater arbitrariness into policy-making and governance processes.

\subsection{Long Term Impacts}
The lenses discussed above reveal that the use of various NLP technologies in the public domain can potentially have long-term impacts on social relationships. These impacts extend beyond mere surface-level economics and have broader implications for knowledge formation, cultural dynamics, the reinforcement of inequity and hierarchy, and more. The instrumental analysis section highlights popular use cases where the decisions made by these systems can be seen as arbitrary to some extent (e.g., systems 2, 3, 26, 27). However, what is often overlooked is the unwarranted trust placed in these often incorrect systems by their human users, who tend to conflate machine-generated results with rigor and may lack the inclination to question them. This behavior has been observed in organizations where officials rely on machine learning for decision-making, without having the incentive or knowledge to critically evaluate the decisions. Unlike human decisions, these machine-generated outcomes are not subject to accountability, petitions, or negotiation. Bad actors within organizations have economic incentives to promote a culture of blind reliance on these systems, and simply acknowledging their imperfections is insufficient without appropriate regulation.

Similarly, the normative analysis highlights the ethical pitfalls of using non-agentic systems to make ethical decisions (especially in medical, legal, and policy contexts). However, what is overlooked is how these systems can effectively erase ethical considerations in the first place when presenting problems to users. For instance, while a lawyer may be alarmed by the idea of NLP systems determining the charges in a crime report, they may not be as concerned if the system merely recommends which documents to read for a case. Yet, even in this seemingly benign use case, the NLP system has already made moral decisions on behalf of the human user, such as determining what factual information is irrelevant. Such erasure of ethical questions is convenient in overburdened legal systems that incentivize quick "solutions" to clear backlogs. Consequently, although individual issues with these systems may be identified, research often overlooks their capability to conveniently transform complex problems within the public sphere. By doing so, these systems effectively remove these problems from public discourse, oversight, and challenge, ultimately maintaining a comfortable status quo.

This research direction leads to a focus on "solving" issues such as bias and transparency as if they were optimization problems, or striving to create "better datasets" in order to engineer away problems stemming from flawed or unsound knowledge, which underlie certain use cases. This distraction prevents researchers from acknowledging the conclusion that, for some of these systems, their public use must be strictly regulated, if not outright prohibited. Fundamentally stochastic systems should not be employed to address problems that require non-stochastic approaches. Problems that are inherently policy or political in nature cannot be solved purely through technical means. Additionally, the research overlooks the significant problem of why NLP systems, despite concerns about flawed datasets, faulty algorithms, potential function creep, and the direct harm to people's rights, exist in the public space. They exist because of the perceived convenience and concrete profitability they offer in obscuring social problems. Our work advocates for future research efforts to understand and challenge the popularity of these systems in order to address these underlying issues.

\section{Limitations and Suggested Research Direction}
The previous analyses underscore the pervasive flaws observed in NLP systems used within the public sphere across various categories. These flaws encompass issues such as unreliable data based on shoddy priors, incomplete and biased datasets, non-transparency in system design, limited user agency, mindless replication of past behaviors through machine learning, and the substitution of stochastic behavior for intelligence. However, it is important to note that these flaws are not solely ethical or design-related; they also stem from inherent limitations within the current state of NLP, especially when applied outside of specific niches and data sources. Moreover, these flaws arise from incorrect and excessive use of NLP systems driven by social and economic factors. By substituting intelligent language and superficial coherence for human accountability, which is crucial in public infrastructure, these problems inevitably arise. The only effective way to address these issues is through rigorous limitations and regulations on the use of NLP systems in the public domain.

While there are instances of genuinely flawed assumptions, such as attempts to attribute emotions to patterns of text, many of these flaws are not purely technical in nature. Additionally, it is essential to recognize that the social gaps these systems aim to "solve" are genuine and cannot be solely rectified by focusing solely on correcting NLP research. The limitations of our paper is that it looks at the various categories of NLP systems but not so much towards the people and organisations which use them for their benefits and despite their flaws.

The ambition of the paper is purely descriptive, offering a taxonomy and in-depth analysis of NLP systems used in the public domain. However, future work is needed to provide concrete policy recommendations tailored to different jurisdictions, political-economic realities, and policy regimes on how these systems should be judiciously used or not used. Furthermore, given that NLP is a socio-technical system, it is imperative to focus on the communities, cultures, and capital involved in the conception and production of these systems. Whether in academia or industry, the design and development of flawed or brittle NLP systems cannot be solely attributed to a lack of ethics training among researchers. Such a perspective would be oversimplified. Instead, we argue that NLP research is both shaped by and implicated in patterns of capital that incentivize a flawed automation of language. Exploring these dynamics will be a focal point of our future work.

To NLP researchers and developers, we urge a broader perspective that considers the societal and economic impact of their work beyond individual ethics and transparency. It is essential to reflect on how their research may shape social and economic relationships. NLP research not only guides the trajectory of the field but also has the potential to influence the generation and distribution of value in society, the evolution of knowledge and culture, and the dynamics of work and wages.

\section{Conclusion}
In recent years, NLP has gained significant popularity in the public domain, thanks to advancements in machine learning research and increased hardware capabilities. These NLP systems have found applications in private usage, market products, and state policy-making, presenting a diverse range of use cases. However, the rapid growth of NLP has outpaced the understanding of scholars studying society and policy, leaving a gap in comprehending the impact of these systems on the "public" sphere. Currently, no comprehensive taxonomy exists that categorizes NLP systems in the public domain and analyzes their social and economic implications.

In this paper, we aim to bridge this gap by offering a broad review of the existing NLP systems in the public domain that have had tangible "real-world" impacts over the past five years. Through thirty illustrative example systems, we analyze the features, limitations, applications, potential misuse, and overall societal implications of these systems. We encourage readers to critically examine whether the research trajectories of these systems are inevitable or influenced by incentives generated by the systems themselves or the political and economic contexts in which they operate. If the current state of affairs is deemed less than ideal, we invite discussions on the future direction of NLP research in its engagement with the "public" sphere.

Our work aims to serve a wide array of scholars, ranging from those seeking a panoramic view of NLP's current public applications to researchers and policymakers investigating potential risks and unintended consequences associated with these systems. Furthermore, we aspire to foster collaboration among NLP researchers, policy experts, economists, and legal scholars to deepen our understanding of the social dimensions inherent in these socio-technical systems.

\section*{Acknowledgements}
The authors would like to thank the anonymous referees for their valuable comments and helpful suggestions. 


\bibliographystyle{IEEEtran}
\bibliography{main_bibfile}

\begin{IEEEbiography}[{\includegraphics[width=1.1in,height=1.25in,clip,keepaspectratio]{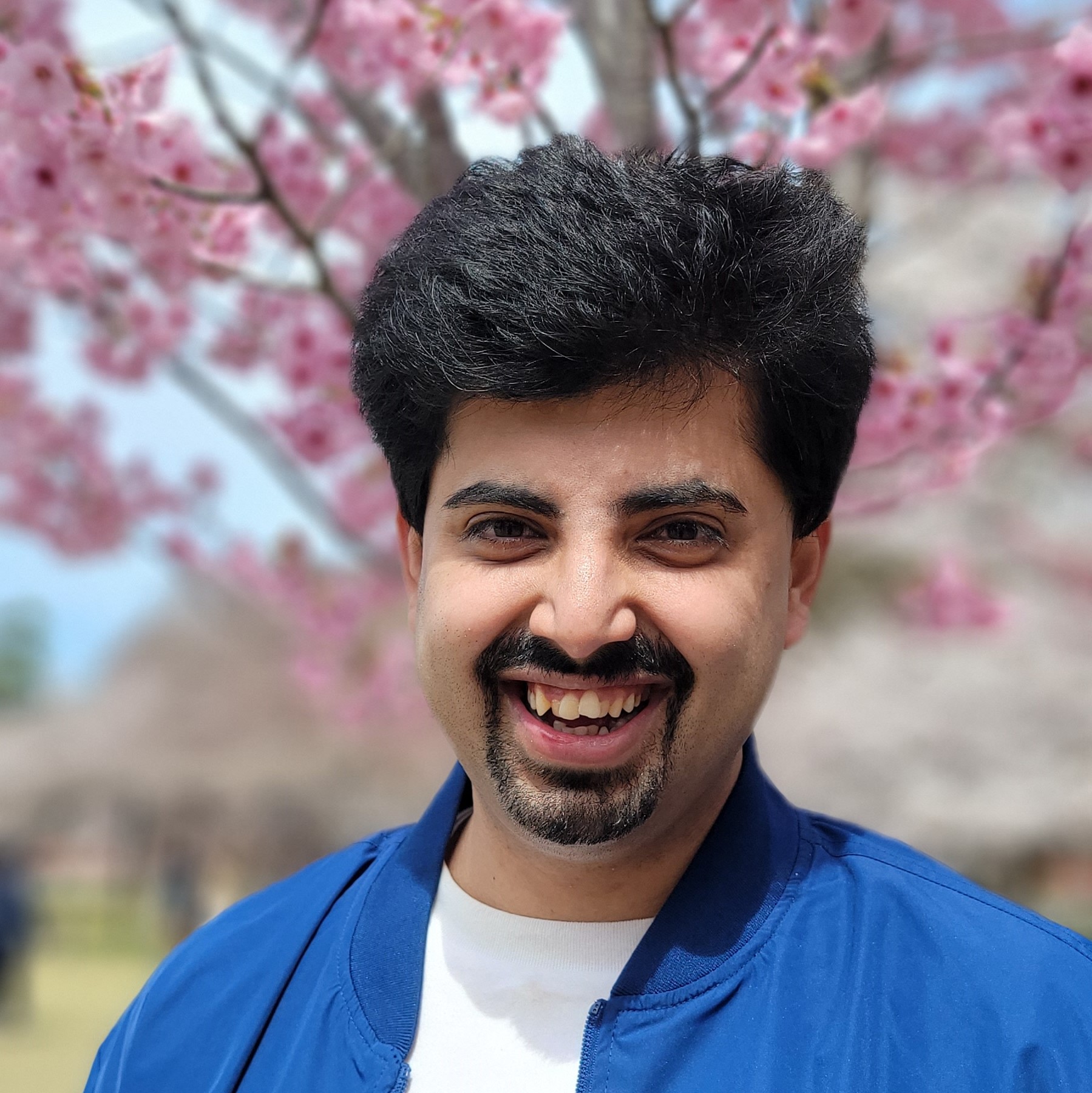}}]{Satyam Mohla}
\
received the Bachelors \& Masters degree in Electrical Engineering from Indian Institute of Technology, Bombay. India. He is currently affiliated with Value Creation Division, Digital Transformation Supervisory Unit, Honda Innovation Lab Tokyo specialising in AI, CASE mobility, digital transformation \& business. He is a Data Science for Social Good Fellow at DFKI, Japan-India Transformative Technology Fellow at Salzburg Global Seminar, Shastri Fellow at Shastri Indo-Canadian Institute and Temasek TfLEaRN Scholar at NTU Singapore.
\end{IEEEbiography}

\begin{IEEEbiography}[{\includegraphics[width=1.1in,height=1.25in,clip,keepaspectratio]{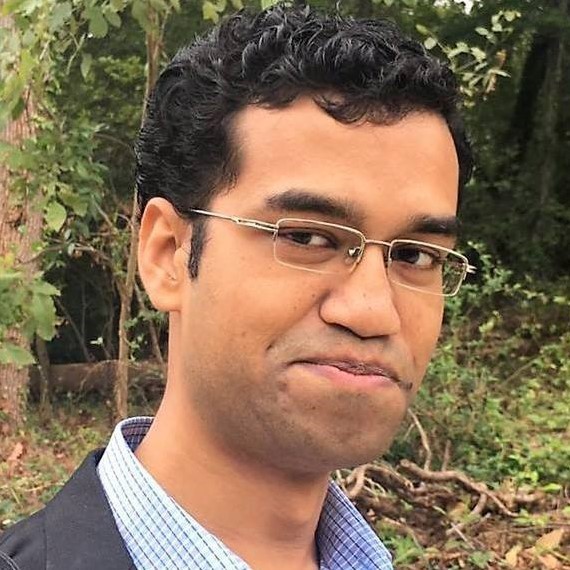}}]{Anupam Guha}
\
received the PhD in Computer Science from University of Maryland in 2017, \& MS in Computer Science from Georgia Tech in 2010. He is an Assistant Professor with the Centre for Policy Studies, Indian Institute of Technology, Bombay. He specialises in working at the intersection of language and vision, and how AI systems work and fail around these problems. 
\end{IEEEbiography}


\end{document}